\documentclass[ draftcls, onecolumn, 12pt]{IEEEtran}   
\usepackage{amsfonts}
\usepackage{graphicx}
\usepackage{epsfig}
\usepackage{amsmath}
\usepackage{amssymb, amsfonts}
\usepackage{latexsym}

\newcommand{\snr}{{\rm SNR}}
\newcommand{\openone}{\leavevmode\hbox{\small1\normalsize\kern-.33em1}}
\newcommand{\field} [1]{\mathbb{#1}}
\newcommand{\boldalpha}{\mbox{\boldmath{$\alpha$}}}
\newcommand{\boldgamma}{\mbox{\boldmath{$\gamma$}}}
\newcommand{\dotleq}{\dot{\leq}}
\newtheorem{theorem}{Theorem}
\newtheorem{proposition}{Proposition}

\title{A Tight Lower Bound to the Outage Probability of Discrete-Input Block-Fading Channels}

\author{Khoa D. Nguyen, Albert Guill\'{e}n i F\`{a}bregas, Lars K. Rasmussen
\thanks{K. D. Nguyen and L. K. Rasmussen are with Institute for Telecommunications Research, University of South Australia, SPRI Building, Mawson Lakes Blvd., Mawson Lakes SA 5095, Australia. e-mail: \tt dangkhoa.nguyen@postgrads.unisa.edu.au, lars.rasmussen@unisa.edu.au.}
\thanks{A. Guill\'{e}n i F\`{a}bregas was with the Institute for Telecommunications Research, University of South Australia, Australia. He is now with the Department of Engineering, University of Cambridge, Trumpington Street, Cambridge CB2 1PZ, UK, e-mail: \tt guillen@ieee.org.}
\thanks{This work has been presented in part at the 2007 IEEE International Symposium on Information Theory, Nice, France, June 2007.}
\thanks{This work has been supported by the Australian Research Council under ARC Grants DP0558861 and RN0459498.}
}

\begin{document}
\maketitle
\begin{abstract}
In this correspondence, we propose a tight lower bound to the outage probability of discrete-input Nakagami-$m$ block-fading channels. The
approach permits an efficient method for numerical evaluation of the bound, providing an additional tool for system
design. The optimal rate-diversity trade-off for the Nakagami-$m$ block-fading channel is also derived and a tight
upper bound is obtained for the optimal coding gain constant.
\end{abstract}


\section{Introduction}
The block-fading channel \cite{OzarowShamaiWyner1994,
BiglieriProakisShamai1998} is a useful channel model for a class of
slowly-varying wireless communication channels. The model is particularly
relevant for delay-constraint applications where channel usage is restricted to
only include a finite number of distinct channel blocks, each subject to
independent flat fading. Frequency-hopping schemes as encountered in the Global
System for Mobile Communication (GSM) and the Enhanced Data GSM Environment
(EDGE), respectively, as well as orthogonal frequency division multiplexing
(OFDM) as encountered in more recently proposed wireless communication systems
standards can conveniently be modeled as block-fading channels. The simplified
model is mathematically tractable, while still capturing the essential features
of the practical slowly-varying fading channels.

In a block-fading channel, a codeword spans a finite number of $B$ independent
fading blocks. As the channel relies on particular realizations of the finite
number of independent fading coefficients, the channel is non-ergodic and
therefore not information stable \cite{VerduHan1994,CaireTariccoBiglieri1999}.
It follows that the Shannon capacity of this channel is zero since there is an
irreducible probability that a given transmission rate $R$ is not supported by
a particular channel realization
\cite{OzarowShamaiWyner1994,BiglieriProakisShamai1998}. This probability is
named the information outage probability. For sufficiently large codes, the
outage probability is the lower bound to the word error rate for any coding
schemes.

Considerable efforts have been dedicated to describing the behavior of the word
error probability and the outage probability for Rayleigh block-fading channels
in the high signal-to-noise ratio (SNR) regime. In particular, analysis based
on worst-case pairwise error probabilities shows that at high SNR the
achievable word error probability of codes $\mathcal{C}$ of rate $R$ (in bits
per channel use) constructed over a signal constellation $\mathcal{X}$ of size
$|\mathcal{X}| = 2^M$ behaves as
\begin{equation}
\lim_{\mathrm{SNR} \rightarrow \infty} -\frac{\log P_e(\mathrm{SNR},R)}{\log \mathrm{SNR}} = d_B(R)
\end{equation}
where
\begin{equation}\label{Singleton}
d_B(R) = 1 + \left\lfloor B \left(1 - \frac{R}{M} \right) \right\rfloor
\end{equation}
is the Singleton bound \cite{Malkamaki1998,MalkamakiLeib1999,KnoppHumblet2000}. More recently, it has been shown
\cite{FabregasCaire2006} that the optimal SNR exponent
\begin{equation}
d^{\star}(R) \triangleq \sup_{\mathcal{C}} \lim_{\mathrm{SNR} \rightarrow \infty} -\frac{\log P_e(\mathrm{SNR},R)}{\log
\mathrm{SNR}}
\end{equation}
is actually given by the Singleton bound (\ref{Singleton}). This establishes
the Singleton bound as the optimal rate-diversity trade-off for transmission
over the Rayleigh block-fading channel with discrete signal constellations.

While these results provide significant insight into code design, the analysis
techniques do not provide explicit tools for the evaluation of the outage
probability; a task which usually requires extensive numerical computations. To
this end, an upper bound to the outage probability of Rayleigh and Rician
block-fading channels is proposed in \cite{Baccarelli1999,
BaccarelliFasano2000}. In this paper, we propose a tight lower bound to the
outage probability which can be efficiently evaluated for the general
Nakagami-$m$ block-fading channel \cite{Nakagami1960}. We show that numerical
evaluation of the proposed bound is very efficient, resulting in significantly
less complex computation as compared to Monte Carlo simulation. We also show
that the optimal rate-diversity trade-off for the Nakagami-$m$ fading case is
given by $d^{\star}(R) = m d_B(R)$ for any $m > 0$, and we obtain an upper
bound to the achievable coding gain for any coding scheme.

The remainder of the correspondence is organized as follows. In Section II, the system
model is described for the Nakagami-$m$ block-fading channel, while Section III
defines the outage probability of this channel. In Section IV, we detail the
proposed lower bound for the outage probability, as well as an efficient method
for the evaluation of the bound. The asymptotic behavior of the outage
probability is investigated in Section V, where the rate-diversity trade-off is
extended to include the Nakagami-$m$ fading statistics. Finally, conclusions
are given in Section VI, while proofs are collected in the Appendices.

The following notation is used in the paper. Sets are denoted by calligraphic
fonts with the complement denoted by superscript $c$. The exponential equality
$g(\xi) \doteq \xi^d$ indicates that $\lim_{\xi \rightarrow \infty} \frac{\log
g(\xi)}{\log \xi} = d$. The exponential inequalities
$\overset{\centerdot}{\leq},\overset{\centerdot}{\geq}$ are similarly defined.
$\openone \{\Psi\}$ is the indicator function for event $\Psi$, $\lceil \xi
\rceil$ $\left(\lfloor \xi \rfloor\right)$ denotes the smallest (largest)
integer greater (smaller) than $\xi$, and $\mathbb{A}^{n}_{+}= \left\{\xi \in
\mathbb{A}^{n}|\xi> 0 \right\}$.

\section{System Model}
\label{sys_mod}

Consider transmission of codewords of length $BL$ coded symbols over a block-fading channel with $B$ blocks. Each block
is an additive white Gaussian noise (AWGN) channel of $L$ channel uses affected by the same flat fading coefficient.
The complex baseband expression for the received signal is
\begin{equation}
\label{channel_model} \mathbf{y}_b = \sqrt{\snr} \; h_b \; \mathbf{x}_b + \mathbf{z}_b, \;\; b= 1, \ldots, B,
\end{equation}
where $\mathbf{y}_b \in \field{C}^L$ is the received signal in block $b$,
$\mathbf{x}_b\in \field{C}^L$ is the portion of the codeword assigned to block
$b$, and $\mathbf{z}_b$ is a noise vector with independent, identically
distributed (i.i.d.) circularly symmetric Gaussian entries
$\sim\mathcal{N}_{\field{C}}(0,1)$. We define $\mathbf{h} = (h_1, \ldots,
h_B)\in \mathbb{C}^B$ as the vector of fading coefficients. The fading
coefficients are assumed i.i.d. from block to block and from codeword to
codeword, as well as being perfectly known to the receiver.

We consider a channel with a discrete input constellation set $\mathcal{X} \subset \field{C}$ of cardinality $2^M$.
Without loss of generality, we assume that $\mathbb{E}[|x|^2]= 1$, where $x \in \mathcal{X}$, and that the fading
coefficients are normalized such that $\mathbb{E}[|h_b|^2]=1$. It follows that SNR is the average signal-to-noise ratio
at the receiver end. Define $\gamma_b \triangleq |h_b|^2$ as the \emph{fading power gain}. Then, the instantaneous
received signal-to-noise ratio at block $b$ is $\gamma_b \snr$.

We consider the case where the fading coefficients follow the general
Nakagami-$m$ distribution \cite{Nakagami1960,SimonAlouini2005}. The probability
density function (pdf) of $|h_b|$ is\footnote{Since the complex coefficients
$h_b$ are perfectly known to the receiver, we can assume phase coherent
detection, and thus, only the amplitude is affected by the fading statistics.}
\begin{equation}
\label{pdf_Nakagami} f_{|h_b|}(\xi) = \frac{2m^m \xi^{2m-1}}{\Gamma(m)} e^{-m\xi^2},
\end{equation}
where $\Gamma(a)$ is the Gamma function $\Gamma(a) = \int_{0}^{\infty} t^{a-1}
\, e^{-t} dt.$ It follows that the fading power gain $\gamma_b$ has the
following pdf
\begin{equation}
\label{pdf_gamma} f_{\gamma_b}(\xi) = \left\{\begin{array}{ll}
\frac{m^m \xi^{m-1}}{\Gamma(m)} e^{-m\xi}, &\xi \geq 0\\
0, &{\rm otherwise,}
\end{array}
\right.
\end{equation}
and cumulative distribution function (cdf)
\begin{equation}
\label{cdf_gamma} F_{\gamma_b}(\xi) =\left\{\begin{array}{ll}
1 - \frac{\Gamma(m , m \xi)}{\Gamma(m)}, & \xi \geq 0\\
0, &{\rm otherwise,}
\end{array}
\right.
\end{equation}
where $\Gamma(a, \xi)$ is the upper incomplete Gamma function $\Gamma(a,\xi) =
\int_{\xi}^{\infty} t^{a-1} \, e^{-t} dt$.

The Nakagami-$m$ distribution represents a large class of fading statistics,
including Rayleigh fading (by setting $m=1$). The distribution also
approximates Rician fading with parameter $K$ (by setting $m= (K+1)^2/(2K+1)$)
\cite{SimonAlouini2005}. Therefore, the proposed analysis for systems with
Nakagami-$m$ fading is a generalization of previous results in the literature.

\section{Mutual Information and Outage Probability}
\label{se:def:outage} The instantaneous input-output mutual information of the block-fading channel with a given
channel realization $\mathbf{h}$ can be expressed as \cite{OzarowShamaiWyner1994}
\[
I (\snr, \mathbf{h}) = \frac{1}{B} \sum_{b=1}^B I_{\rm AWGN}(\gamma_b \snr),
\]
where $I_{\rm AWGN}(\rho)$ is the input-output mutual information of an AWGN
channel with SNR $\rho$. $I(\snr, \mathbf{h})$ is the input-output mutual
information of a set of $B$ non-interfering parallel channels, each of which is
used only for a fraction $\frac{1}{B}$ of the time. When the input signal set
$\mathcal{X}$ is discrete, the mutual information $I_{\rm AWGN}(\rho)$ is given
by
\begin{equation}
I_{\rm AWGN}(\rho) = M - 2^{-M}\sum_{x \in \mathcal{X}} \mathbb{E}\left[\log_2 \left(\sum_{x' \in \mathcal{X}}
e^{-|\sqrt{\rho}(x-x')+Z|^2+ |Z|^2} \right)\right],
\end{equation}
where the expectation over $Z \sim \mathcal{N}_{\field{C}}(0, 1)$ can be efficiently computed using the
Gauss-Hermite quadrature rules \cite{AbramowitzStegun1964}.

Transmission at rate $R$ over the channel in \eqref{channel_model} is considered to be in outage whenever
\[
\frac{1}{B} \sum_{b=1}^B I_{\rm AWGN}(\gamma_b {\rm SNR}) < R.
\]
The corresponding outage probability is given by
\begin{equation}
\label{outage_form} P_{\rm{out}} (\snr, R)= \Pr \left(\frac{1}{B} \sum_{b=1}^B I_{\rm AWGN}(\gamma_b \snr) < R\right).
\end{equation}

\section{Lower Bound to the Outage Probability}
\label{lowerbound_outage} In general, when the channel has a discrete input
constellation, evaluation of the outage probability in (\ref{outage_form}) is
complicated since a closed form expression for $I_{\rm AWGN}(\rho)$ is not
known. Typically, $P_{\rm out}({\rm SNR}, R)$ is instead evaluated through
Monte Carlo simulations\footnote{Even if the inputs to the channel are Gaussian,
for which $I_{\rm AWGN}(\gamma_b {\rm SNR}) = \log_2(1+\gamma_b {\rm SNR})$, a
closed form expression for the outage probability is not known.}, which are computationally demanding for high SNR. In this
section, we propose a lower bound to the outage probability with discrete
inputs, which can be efficiently computed for any SNR.

The maximum input-output mutual information for a channel with input signal
constellation $\mathcal{X}$ of size $|\mathcal{X}| = 2^M$ is always upper
bounded by $M$. Furthermore, the input-output mutual information of the channel
can also be upper bounded by that of the channel with Gaussian input.
Therefore, for any realization of $\boldgamma$, $I_{\rm AWGN}(\gamma_b \snr),
b=1, \ldots, B$ is upper bounded by\footnote{Superscripts $u$ and $\ell$ will
denote upper and lower bounds respectively.}
\begin{eqnarray}
\label{upperbound_mutual}
I_{\rm AWGN} ^u (\gamma_b \snr) &\triangleq& \min\{M, \log_2(1+\gamma_b \snr)\} \\
\label{form_Iu} &=& \left\{\begin{array}{ll}
\log_2(1+\gamma_b \snr), &\gamma_b \leq \frac{2^M-1}{\snr}\\
M, &\mbox{otherwise}
\end{array}\right. \notag\\
\label{form_Iu_S} &=&\left\{\begin{array}{ll}
\log_2(1+ \gamma_b \snr), & b \in \mathcal{S}^c\\
M,& b \in \mathcal{S},
\end{array}\right.
\end{eqnarray}
where $\mathcal{S} = \left\{b \in \{1, 2, \ldots, B\}: \gamma_b > \frac{2^M-1}{\snr}\right\}$ and $\mathcal{S}^c$
denotes its complement.

Let $|\mathcal{S}|$ be the cardinality of $\mathcal{S}$. Since $\gamma_b, \;
b=1, \ldots, B,$ are independent random variables, $|\mathcal{S}|$ is a
binomially distributed random variable with success rate $p \triangleq
\Pr\left(\gamma_b > \frac{2^M-1}{\snr}\right)$. Hence,
\begin{equation}
\Pr(|\mathcal{S}|= t) = \binom{B}{t} p^t (1-p)^{B-t}, \;\;\; t= 1, 2, \ldots,
B,
\end{equation}
where
\begin{eqnarray}
p
&=& 1- F_{\gamma_b}\left(\frac{2^M-1}{\snr}\right) \notag\\
\label{succ_rate}&=& \frac{\Gamma \left(m, m \frac{2^M-1}{\snr}\right)}{\Gamma(m)}.
\end{eqnarray}
Using the upper bound of mutual information in (\ref{upperbound_mutual}) and
(\ref{form_Iu_S}), we lower bound $P_{\rm out}({\rm SNR}, R)$ as
\begin{align}
P_{\rm out}^{\ell}(\snr, R) &\triangleq \Pr \left(\frac{1}{B} \sum_{b=1}^B I_{\rm AWGN} ^u (\gamma_b {\rm SNR}) < R \right)
\label{eq:lower_bound}\\
&= \Pr\left( \sum_{b \in \mathcal{S}} I_{\rm AWGN}^u  (\gamma_b \snr) + \sum_{b \in \mathcal{S}^c} I_{\rm AWGN}^u(\gamma_b \snr)< BR\right)\\
&= \Pr\left(|\mathcal{S}|M + \sum_{b \in \mathcal{S}^c} \log_2(1+\gamma_b \snr) < BR
\right).\label{eq:lower_bound_cardS}
\end{align}
Since $\gamma_b, \; b= 1, \ldots, B$ are i.i.d. random variables, $\sum_{b \in
\mathcal{S}^c} \log_2(1+\gamma_b \snr)$ is the summation of $|\mathcal{S}^c| =
B- |\mathcal{S}|$ i.i.d. random variables. Each random variable inside the
summation is given by $\log_2(1+ \gamma_b \snr)$ conditioned on $b \in
\mathcal{S}^c$, or equivalently on the event $\mathcal{E}$, where $\mathcal{E}$
is defined as
\begin{equation}
\label{eq:event:E}\mathcal{E} \triangleq \left\{\gamma_b: \gamma_b \leq \frac{2^M-1}{\snr}\right\}.
\end{equation}
Denote $A_b$ as the random variable $\log_2(1+\gamma_b \snr)$ conditioned on
$\mathcal{E}$. Then, the distribution of $A_b$ is given by the following
proposition.
\begin{proposition}
\label{prop:dist:Ab} Assume $\gamma_b$ is a random variable whose distribution is given by \eqref{pdf_gamma}. Denote
$A_b$ as the random variable $\log_2(1+\gamma_b \snr)$ conditioned on the event $\mathcal{E}$ given in
\eqref{eq:event:E}. The distribution of $A_b$ is then given by
\begin{equation}
\label{pdfA_b} f_{A_b}(\xi) = \left\{ \begin{array}{ll}
\frac{f_{\gamma_b}\left(\frac{2^\xi-1}{\snr}\right)}{F_{\gamma_b}\left(\frac{2^M-1}{\snr}\right)} \frac{2^\xi
\log(2)}{\snr}, & 0 \leq \xi \leq M \\
0, & \rm{otherwise}.
\end{array}\right.
\end{equation}
\end {proposition}
\begin{proof}
See Appendix \ref{app:distAb}
\end{proof}
Therefore, denoting $A_k, k=1, \ldots, |\mathcal{S}^c|$, as the
$B-|\mathcal{S}|$ independent random variables that follow the distribution
given in (\ref{pdfA_b}), we can write (\ref{eq:lower_bound_cardS}) as
\begin{equation}
P_{\rm out}^{\ell}(\snr, R)= \Pr\left(|\mathcal{S}|M + \sum_{k=1}^{B-|\mathcal{S}|} A_k< BR \right).
\end{equation}
By conditioning on $|\mathcal{S}|$, we can express $P_{\rm out} ^{\ell} ({\rm SNR}, R)$ as
\begin{align}
P_{\rm out} ^{\ell} ({\rm SNR}, R) &= \sum_{t=0}^B\Pr \left( \left.\sum_{k=1}^{B-|\mathcal{S}|}A_k < BR-|\mathcal{S}|M
\right| |\mathcal{S}|=t\right) \Pr(|\mathcal{S}|=t)\\
&= \sum_{t=0}^B\Pr \left(\sum_{k=1}^{B-t}A_k < BR-tM \right) \Pr(|\mathcal{S}|=t).\label{eq:out_lb_binom}
\end{align}
From the distribution in (\ref{pdfA_b}), note that $\Pr(A_k \leq 0)= 0$.
Therefore, for any $t$ such that $BR-tM \leq 0$, or equivalently for all $t
\geq \left\lceil\frac{BR}{M}\right\rceil$, the corresponding probability is
zero. Hence, we can rewrite \eqref{eq:out_lb_binom} as
\begin{equation}
\label{eq:se4:last_conded}P_{\rm out} ^{\ell} ({\rm SNR}, R)= \sum_{t=0}^{\left
\lceil \frac{BR}{M}\right\rceil-1} \Pr \left(\sum_{k=1}^{B-t} A_k< BR-tM
\right) \Pr(|\mathcal{S}|=t).
\end{equation}
If we now define the random variable $Y_t \triangleq \sum_{k=1}^{B-t} A_k$, we can write
\begin{equation}
\label{P_lout_Y} P_{\rm out}^{\ell} ({\snr}, R)=
\sum_{t=0}^{\left\lceil\frac{BR}{M}\right\rceil-1}F_{Y_t}(BR-tM)
\binom{B}{t}p^t(1-p)^{B-t},
\end{equation}
where $F_{Y_t}(\xi)$ is the cdf of $Y_t$.

Since $A_k, k=1, \ldots,B-t$ are independent random variables, the pdf of
${Y_t}$ can be evaluated by performing $B-t$ convolutions of $f_{A_b}(\xi)$.
Numerically, this convolution can be efficiently computed in the frequency
domain using fast Fourier transform (FFT) techniques
\cite{ProakisManolakis1992}. With this method, we can efficiently evaluate the
cdf of $Y_t, \ F_{Y_t}(\xi)$, and therefore we can also efficiently evaluate
$P_{\rm out}^{\ell}(\snr, R)$ in \eqref{P_lout_Y}. The evaluation of
\eqref{P_lout_Y} is significantly faster than evaluating $P_{\rm out}(\snr, R)$
in \eqref{outage_form} using Monte Carlo simulation techniques.

\begin{figure}[htbp]
\begin{center}
  \includegraphics[width=1 \columnwidth]{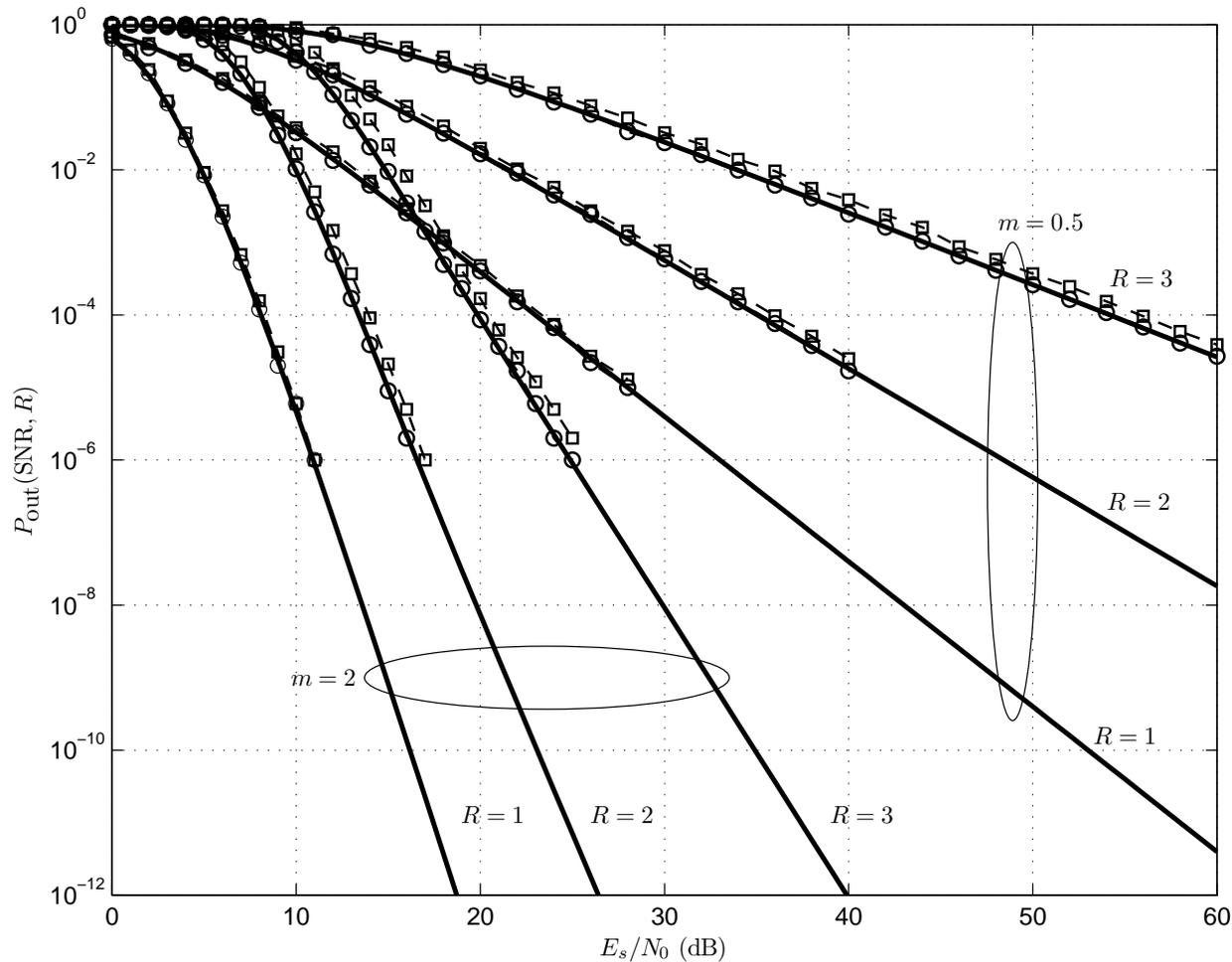}\\
  \caption{Outage probability of Nakagami-$m$ block-fading channels with $B=4, M=4$, $m=0.5$ and $m=2$. The thick solid lines correspond to the lower
  bound ({\ref{P_lout_Y}}), thin dashed lines with circles denote the simulation of \eqref{eq:lower_bound} and thin dashed lines with
  squares denote the simulation of \eqref{outage_form} with $16$-QAM modulation.}
  \label{fig_two}
\end{center}
\end{figure}

\begin{figure}[htbp]
\begin{center}
   \includegraphics[width = 1 \columnwidth]{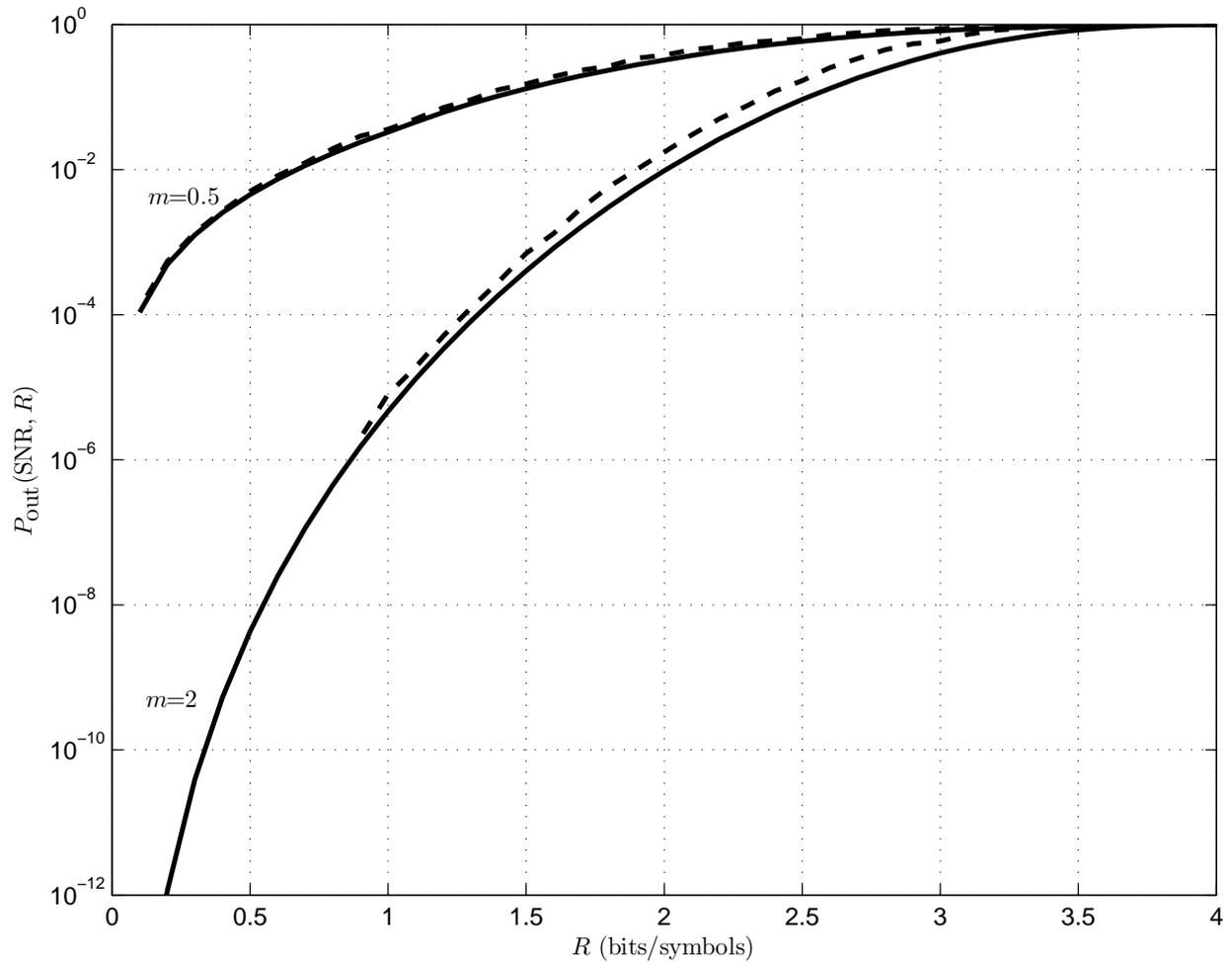}\\
  \caption{Outage probability for the of Nakagami-$m$ block-fading channels with $B=4, M=4, \snr = 10{\rm dB}$, $m=0.5$ and $m=2$. The
  solid lines correspond to the lower bound \eqref{P_lout_Y}. The dashed lines denote the simulation of \eqref{outage_form} with 16-QAM modulation.}
  \label{fig:outage:vs:rate}
\end{center}
\end{figure}

Numerical results for Nakagami-$m$ block-fading channels with $B=4$, $M=4$, $m=0.5$ and
$m=2$ are given in Figure \ref{fig_two}. The transmission rates considered are
$R=1, 2, 3$ bits per channel use, which correspond to Singleton bounds
$d_B(R)= 4, 3, 2$, respectively. The figure shows the simulation and
analytical curves of the lower bound to the outage probability of the channel
based on \eqref{eq:lower_bound} and \eqref{P_lout_Y}, respectively, together
with the 16-QAM outage simulation curve based on \eqref{outage_form}. We
observe that the analytical curves coincide with the corresponding lower bound
simulation curves. The analytical curves give a tight lower bound to the 16-QAM
outage curve. Note that the bound is very tight for the important case of
$R=1$, which, from the Singleton bound expression in \eqref{Singleton}, is the
largest rate that can be achieved with full diversity. Figure
\ref{fig:outage:vs:rate} provides a plot of the outage probability of the same
channels as a function of the code rate $R$ at $\snr = 10$dB, illustrating the
validity of the bound over a wide range of transmission rates. Further
simulations show that these observations are valid for a wide range of channel
parameters. We also observe from Figure \ref{fig_two} that the slope of each
curve is $m d_B(R)$, representing the SNR exponent of the outage probability.
In the following section, we rigorously prove that the optimal SNR-exponent
over the channel is
\begin{equation}
d^{\star} (R)= m d_B(R).
\end{equation}
In proving this result, we characterize not only the SNR-exponent but also the
asymptotic coding gain.
%

\section{Asymptotic Behavior}
\label{asymptotic} Using \eqref{P_lout_Y} and the analysis techniques from
\cite{FabregasCaire2006}, we obtain the following result for the asymptotic
diversity of Nakagami-$m$ block-fading channels, for all $m>0$.
\begin{proposition}
\label{prop_converse} Assume transmission over the block-fading channel as
defined in (\ref{channel_model}) with input signal constellation size $2^M$.
Assume further that the fading power gain $\gamma_b$ is a random variable whose
distribution is given by (\ref{pdf_gamma}). In this case, the lower bound on
$P_{\rm out}(\snr, R)$ given in \eqref{P_lout_Y} can asymptotically be
expressed as
\begin{equation}
\label{eq:asym:Plout:prop} P^{\ell}_{\rm out}(\snr, R) \doteq
\mathcal{K}_{\ell} \snr^{-m d_B(R)},
\end{equation}
where $d_B(R)$ is the Singleton bound given in (\ref{Singleton}). Furthermore,
$\mathcal{K}_{\ell}$ is a constant independent of $\snr$ given by
\begin{equation}
\label{eq:Kl:prop:Plout}\mathcal{K}_{\ell}=
F_{\overline{Y}_{B-d_B(R)}}\Big(BR-(B-d_B(R))M\Big) \binom{B}{B-d_B(R)}
\frac{(m (2^M-1))^{md_B(R)}}{(m \Gamma(m))^{d_B(R)}},
\end{equation}
where
\begin{equation}
F_{\overline{Y}_t} (\xi) = \lim_{\snr \to \infty} F_{Y_t}(\xi)
\end{equation}
\end{proposition}
\begin{proof}
See Appendix \ref{app:Asymp:Plout}.
\end{proof}

This proposition not only shows that the SNR exponent of the outage probability
is upper bounded by $m d_B(R)$ but also gives the asymptotic constant
$\mathcal{K}_{\ell}$ of $P^{\ell}_{\rm out}(\snr, R)$. This is indeed useful
for code design since it gives an upper bound for the coding gain achieved by
any coding scheme. At the same time, together with the expression of $P_{\rm
out}^{\ell}(\snr, R)$ given in \eqref{P_lout_Y}, it gives a more specific
characterization of the outage probability, indicating the word error
probability (or SNR) region where asymptotic analysis is valid.

\begin{figure}[htbp]
\begin{center}
   \includegraphics[width=1 \columnwidth]{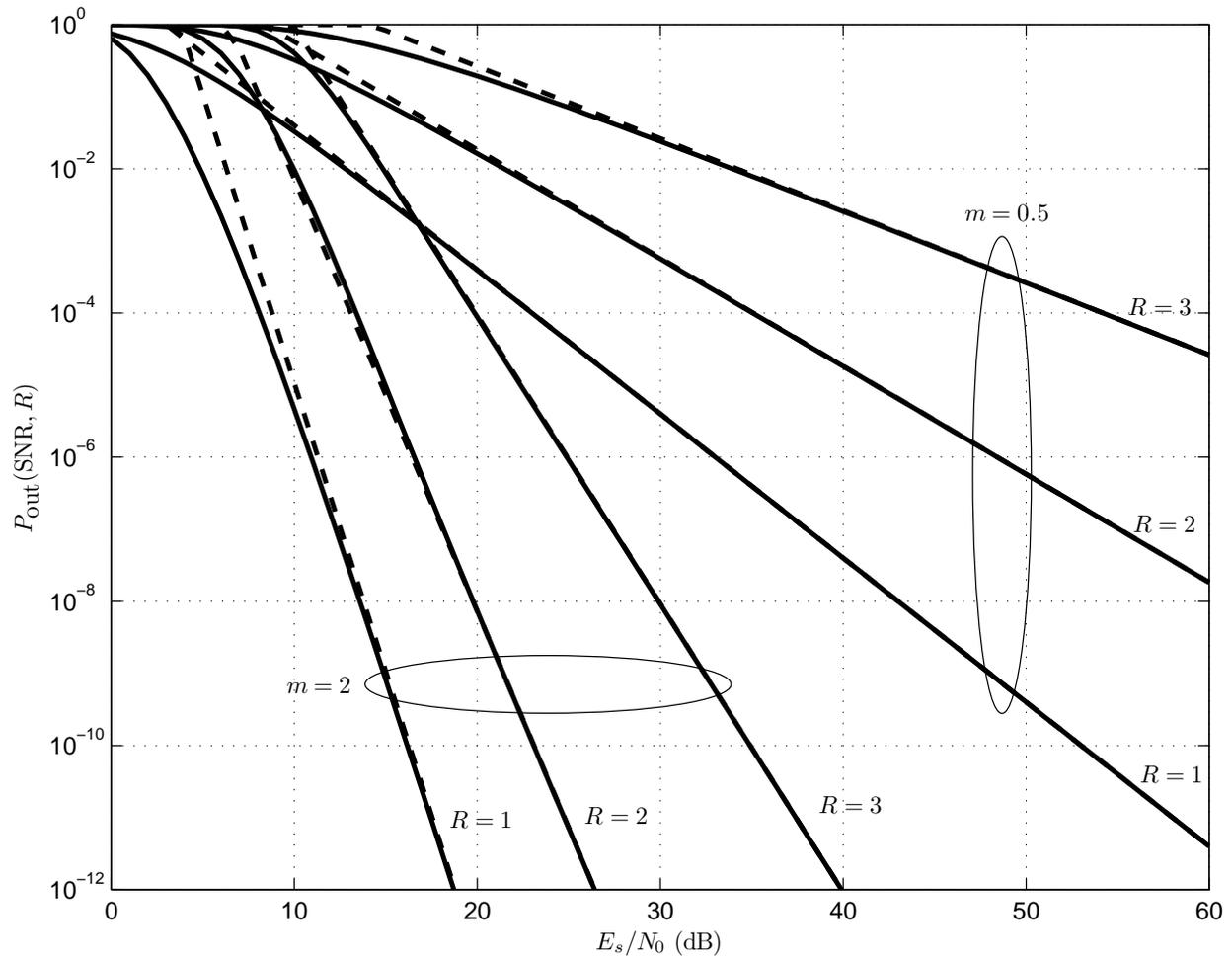}\\
  \caption{Outage probability of Nakagami-$m$ block-fading channels with $B=4, M=4$, $m=0.5$ and $m=2$. The solid lines correspond to
  the lower bound ({\ref{P_lout_Y}}) and the dashed lines to its asymptotic expression given in \eqref{eq:asym:Plout:prop}
  using $\mathcal{K}_{\ell}$ in \eqref{eq:Kl:prop:Plout}.}
  \label{fig_asym_two}
\end{center}
\end{figure}

The lower bound to the outage probability and the asymptotic term given in
\eqref{eq:asym:Plout:prop} are illustrated in Figure \ref{fig_asym_two}. The
same set of parameters as in Figure \ref{fig_two} has been chosen, namely
$B=4,M=4, m=2$ and $R=1,2,3$.

So far, we have shown that $d^\star(R)\leq m d_B(R)$. To prove the optimality
of the SNR-exponent $m d_B(R)$, we need to prove the achievability result given
in the next proposition.
\begin{proposition}
\label{prop_achieve} Assume transmission with random codes of rate $R$ and
block length $L(\snr)$ satisfying
\begin{equation}
\label{eq:lambda:blocklength} \lim_{\snr\to\infty} \frac{L(\snr)}{\log(\snr)} = \lambda
\end{equation}
over a block-fading channel as defined in (\ref{channel_model}) with input
signal constellation size $2^M$. Further assume that the fading power gain
$\gamma_b$ is a random variable whose distribution is given by
(\ref{pdf_gamma}). In this case, the SNR-exponent is lower bounded by
\begin{equation}
\label{eq:diversity_achieve}
d^{(r)}(R) \geq \left\{\begin{array}{ll}
\lambda B M  \log (2) (1- \frac{R}{M}), &\lambda < \frac{m}{M \log(2)}\\
m(d_B(R)-1) + \min\left\{m, \lambda M \log(2) \left( B
\left(1-\frac{R}{M}\right) - d_B(R)+1\right)\right\},&\lambda \geq \frac{m}{M
\log(2)}.
\end{array}\right.
\end{equation}
\end{proposition}
\begin{proof}
See Appendix \ref{app:diversity:achive}.
\end{proof}
The preceding propositions lead to the following theorem.
\begin{theorem}
\label{diversity_achieve_theorem} Assume transmission over a block-fading
channel as defined in (\ref{channel_model}) with input constellation size
$2^M$. Further assume that the fading power gain $\gamma_b$ is a random
variable whose distribution is given by (\ref{pdf_gamma}). In this case, the
optimal SNR-exponent is given by
\begin{equation}
d^{\star}(R) = m d_B(R)
\end{equation}
for all $R, M$ where $B\left(1-\frac{R}{M}\right)$ is not an integer.
\end{theorem}
\begin{proof}
See Appendix \ref{app:diversity:optimal}.
\end{proof}
As remarked in Appendix \ref{app:diversity:optimal}, Theorem
\ref{diversity_achieve_theorem} can be proved using the methods proposed in
\cite{FabregasCaire2006}. However, with the proof proposed here, Propositions
\ref{prop_converse} and \ref{prop_achieve} provide additional information. In
particular, Proposition \ref{prop_converse} provides an upper bound on the
coding gain $\mathcal{K}_{\ell}$, and Proposition \ref{prop_achieve} provides
an extension for the SNR-exponent of random codes with finite block length in
\cite{FabregasCaire2006} to a more general fading distribution.

\begin{figure}[htbp]
\begin{center}
  \includegraphics[width = 1 \columnwidth]{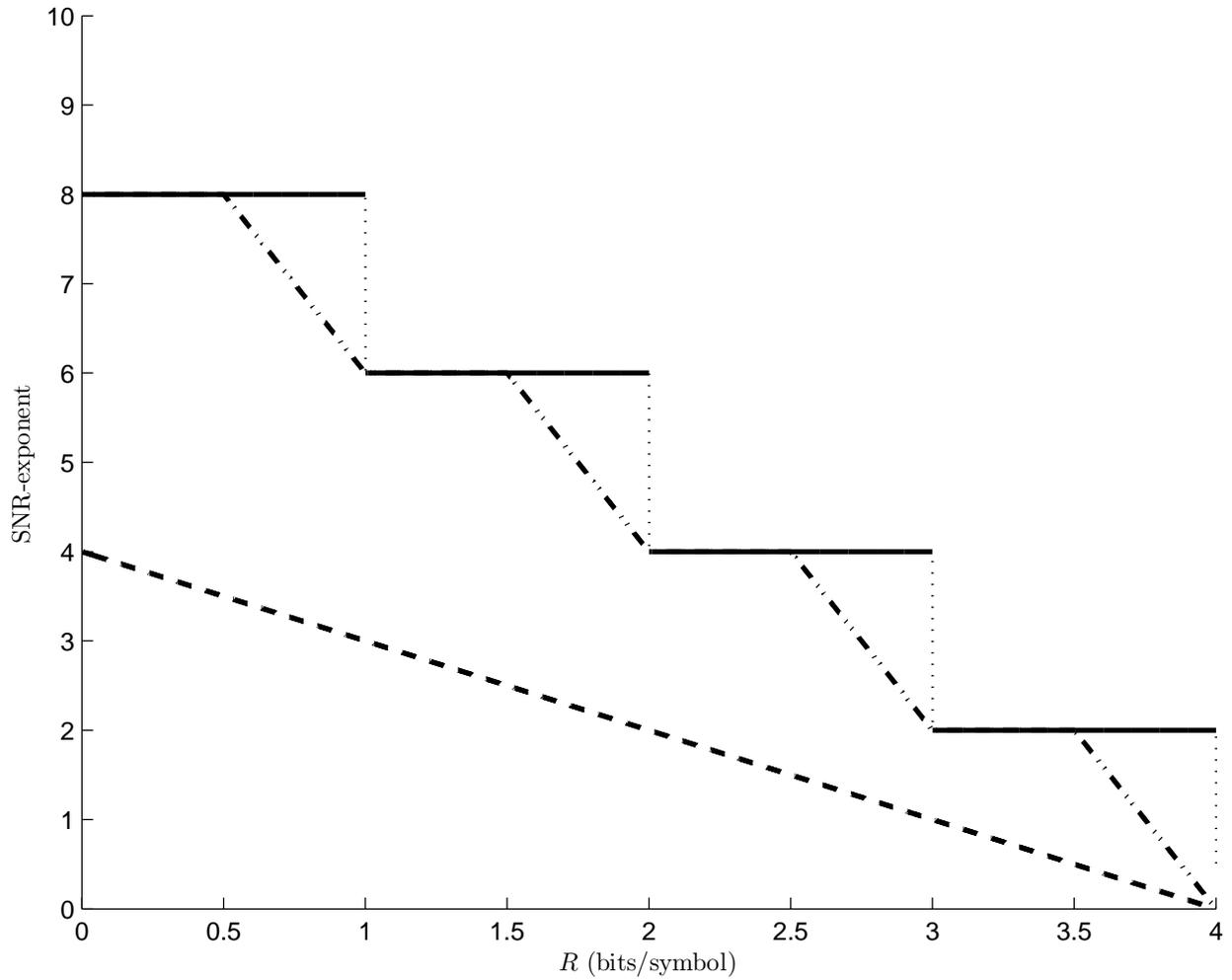}\\
  \caption{Optimal and random coding SNR-exponent for Nakagami-$m$ block-fading channels with $m=2, B=4, M=4$.
  The solid line corresponds to $md_B(R)$, dashed-dotted line and dashed line denote the random coding exponent with
  $\lambda M \log(2) = 2m$ and $\lambda M \log(2) = \frac{m}{2}$ respectively.}
  \label{fig_div_two}
\end{center}
\end{figure}

The diversity of random codes for block-fading channels with $B =4$, $M=4$ and
$m=2$ is illustrated in Figure \ref{fig_div_two}. Random codes with block
length satisfying $\lambda = \frac{2m}{M \log(2)}$ and $\lambda = \frac{m}{2M
\log(2)}$ are considered, where $\lambda$ is defined in
\eqref{eq:lambda:blocklength}. We observe that the SNR-exponent is always upper
bounded by $m d_B(R)$. Except for points of discontinuity of $d_B(R)$, the
upper bound can be achieved by increasing $\lambda$ since $d^{(r)}(R)$ and $m
d_B(R)$ will coincide over larger ranges of $R$.
\section{Conclusion}
\label{Concluding} In this correspondence, we have proposed a tight lower bound to the
outage probability of discrete-input block-fading channels with Nakagami-$m$ fading
statistics. The lower bound can be computed efficiently and is therefore useful
for system design and analysis. We show that the optimal rate-diversity trade-off for
Nakagami-$m$ block-fading channels is given by $m$ times the Singleton bound. We also obtain an upper bound for the achievable
coding gain, which is useful for code design.
\newpage
\appendices
\section{Distribution and Properties of $A_b$}
\label{app:distAb}
\paragraph*{Proposition \ref{prop:dist:Ab}}
Assume $\gamma_b$ is a random variable whose distribution is given by
\eqref{pdf_gamma}. Denote $A_b$ the random variable $\log_2(1+\gamma_b \snr)$
conditioned on the event $\mathcal{E}$ described in \eqref{eq:event:E}. The
distribution of $A_b$, is given by
\begin{equation}
\label{pdfA_b_app} f_{A_b}(\xi) = \left\{ \begin{array}{ll} \frac{f_{\gamma_b}
\left(\frac{2^\xi-1}{\snr}\right)}{F_{\gamma_b}\left(\frac{2^M-1}{\snr}\right)}
\frac{2^\xi
\log(2)}{\snr}, & 0 \leq \xi \leq M \\
0, & \rm{otherwise}.
\end{array}\right.
\end{equation}

\begin{proof}
The cdf of $A_b$ is given by
\begin{align}
F_{A_b}(\xi) &= \Pr(\log_2(1+\gamma_b \snr)<\xi|\mathcal{E})\\
&=\Pr\left(\log_2(1+\gamma_b \snr)< \xi\left| \gamma_b \leq
\frac{2^M-1}{\snr}\right.\right).
\end{align}
Applying Bayes' rule, we obtain
\begin{equation}
\label{fA_after_bay} F_{A_b}(\xi) = \frac{\Pr\left(\gamma_b <
\frac{2^\xi-1}{\snr}, \gamma_b \leq \frac{2^M-1}{\snr}\right)}{\Pr
\left(\gamma_b \leq \frac{2^M-1}{\snr}\right)}.
\end{equation}
If $\xi \leq M$ then $2^\xi-1 \leq 2^M-1$ and therefore,
\begin{eqnarray}
 \Pr \left(\gamma_b < \frac{2^\xi-1}{\snr},  \gamma_b \leq \frac{2^M-1}{\snr}\right) &=& \Pr\left(\gamma_b
<\frac{2^\xi-1}{\snr}\right)\\
\label{fA_xlM} &=& F_{\gamma_b}\left(\frac{2^\xi-1}{\snr}\right).
\end{eqnarray}
Otherwise, if $\xi> M$,
\begin{eqnarray}
 \Pr \left(\gamma_b < \frac{2^\xi-1}{\snr}, \gamma_b \leq \frac{2^M-1}{\snr}\right) &=& \Pr\left(\gamma_b
\leq \frac{2^M-1}{\snr}\right)\\
\label{fA_xgM}&=& F_{\gamma_b}\left(\frac{2^M-1}{\snr}\right).
\end{eqnarray}
By inserting (\ref{fA_xlM}) and (\ref{fA_xgM}) into (\ref{fA_after_bay}), we
finally have that
\begin{equation}
\label{eq:appI:cdfA_b}F_{A_b}(\xi) = \left\{
\begin{array}{ll}
\frac{F_{\gamma_b}\left(\frac{2^\xi-1}{\snr}\right)}{F_{\gamma_b}\left(\frac{2^M-1}{\snr}\right)}, & \xi \leq M  \\
1, & \mbox{otherwise}.
\end{array}
\right.
\end{equation}
Now differentiate $F_{A_b}(\xi)$ in (\ref{eq:appI:cdfA_b}) with respect to
$\xi$, noting that $\frac{d}{d\xi}F_{A_b}(\xi)= f_{A_b}(\xi)$ and
$\frac{d}{d\xi}F_{\gamma_b}(\xi)= f_{\gamma_b}(\xi)$, we obtain (\ref{pdfA_b}).
\end{proof}

\begin{proposition}
\label{prop:dist:Ab:asym} Assume $\gamma_b$ is a random variable whose
distribution is given by \eqref{pdf_gamma}. Assume $A_b$ is a random variable
as defined in Proposition \ref{prop:dist:Ab}. Asymptotically, the distribution
of $A_b$ is independent of $\snr$ and is given by
\begin{equation}
\label{asymp_pdfA_b} f_{A_b}(\xi) \doteq f_{\overline{A}_b}(\xi) \triangleq
\left\{\begin{array}{ll}
\frac{m(2^\xi-1)^{m-1}2^\xi \log(2)}{(2^M-1)^m}, & \xi \leq  M, \\
0, &\rm{otherwise}.
\end{array}\right.
\end{equation}
\end {proposition}
\begin{proof}
From (\ref{pdf_gamma}) and Taylor series expansion, we have
\begin{eqnarray}
f_{\gamma_b}\left(\frac{2^\xi-1}{\snr}\right) &=& \frac{m^m
\left(\frac{2^\xi-1}{\snr}\right)^{m-1}}{\Gamma(m)}e^{-m
\frac{2^\xi-1}{\snr}} \notag\\
\label{asymp_pdf_gamma}&\doteq& \frac{m^m (2^\xi-1)^{m-1}}{\Gamma(m)}
\snr^{-(m-1)}.
\end{eqnarray}
Similarly, from (\ref{cdf_gamma}), we have
\begin{eqnarray}
F_{\gamma_b}\left(\frac{2^M-1}{\snr}\right)  &=& 1 - \frac{\Gamma\left(m, m \frac{2^M-1}{\snr}\right)}{\Gamma(m)} \notag\\
&\doteq& 1-\frac{\Gamma(m)- \frac{1}{m}\left(m \frac{2^M-1}{\snr}\right)^m}{\Gamma(m)} \notag\\
\label{asymp_cdf_gamma}&\doteq& \frac{m^m(2^M-1)^m}{m \Gamma(m)} \snr^{-m}.
\end{eqnarray}
Inserting (\ref{asymp_pdf_gamma}) and (\ref{asymp_cdf_gamma}) into
(\ref{pdfA_b_app}), we obtain (\ref{asymp_pdfA_b}).
\end{proof}
\newpage
\section{Proof of Proposition \ref{prop_converse}}
\label{app:Asymp:Plout} Define $\overline{A}_k$ as a random variable described
by the distribution function $f_{\overline{A}_b}(\xi)$ given in
\eqref{asymp_pdfA_b}. Further define $F_{\overline{Y}_t}(\xi)$ as the cdf of
$\overline{Y}_t \triangleq \sum_{k=1}^{B-t} \overline{A}_k$. According to
Proposition \ref{prop:dist:Ab:asym}, $f_{A_b}(\xi) \doteq
f_{\overline{A}_b}(\xi)$, and therefore, $F_{Y_t}(\xi) \doteq
F_{\overline{Y}_t}(\xi)$. In addition, Taylor expansion of (\ref{succ_rate})
gives
\begin{eqnarray}
\label{asymp_p}p &\doteq& \frac{\Gamma(m)- \frac{1}{m}(m \frac{2^M-1}{\snr})^m}{\Gamma(m)} \doteq 1, \\
\label{asymp_1-p}1-p &\doteq& \frac{m^m (2^M-1)^m}{m \Gamma(m)} \snr^{-m}.
\end{eqnarray}
Since the asymptotic expressions for $p, 1-p$ and $F_{Y_t}(\xi)$ are finite
and non-zero, the asymptotic behavior of $P^{\ell}_{\rm out}({\rm SNR}, R)$ in
(\ref{P_lout_Y}) is found by replacing $F_{Y_t}(\xi)$ with
$F_{\overline{Y}_t}(\xi)$, and replacing $p, 1-p$ with their corresponding
asymptotic value in \eqref{asymp_p} and \eqref{asymp_1-p}. It follows that
\begin{equation}
\label{asymp_Plout} P^{\ell}_{\rm out} ({\rm SNR}, R) \doteq
\sum_{t=0}^{\left\lceil\frac{BR}{M}\right\rceil-1} F_{\overline{Y}_t}(BR-tM)
\binom{B}{t} \left(\frac{m^m (2^M-1)^m}{m \Gamma(m)}\right) ^{B-t}
\snr^{-m(B-t)}.
\end{equation}
Since $f_{\overline{A}_b}(\xi)$ is independent of $\snr$,
$F_{\overline{Y}_t}(\xi)$ is also independent of $\snr$. Therefore, the term
with minimum $m(B-t)$ dominates the expression in (\ref{asymp_Plout}). The
dominating term corresponds to
\begin{equation}
t = \left\lceil \frac{BR}{M}\right\rceil - 1,
\end{equation}
and thus
\begin{equation}
B-t = 1 + \left\lfloor B\left(1-\frac{R}{M}\right)\right\rfloor = d_B(R),
\end{equation}
which is precisely the Singleton bound. Therefore, we write the asymptotic
behavior for (\ref{P_lout_Y}) as
\begin{equation}
\label{asymp_Plout_fined} P^{\ell}_{\rm out}({\rm SNR}, R) \doteq
\mathcal{K}_{\ell} \,\snr^{-md_B(R)},
\end{equation}
where
\begin{equation}
\label{eq:Kl:app:proof:Plout}\mathcal{K}_{\ell}=
F_{\overline{Y}_{B-d_B(R)}}\Big(BR-(B-d_B(R))M\Big) \binom{B}{B-d_B(R)}
\frac{(m (2^M-1))^{md_B(R)}}{(m \Gamma(m))^{d_B(R)}}
\end{equation}
is independent of $\snr$.
\newpage
\section{Proof of Proposition \ref{prop_achieve}}
\label{app:diversity:achive} The proof follows the same lines as in
\cite{FabregasCaire2006} with the generalization of Rayleigh fading statistic
to Nakagami-$m$ fading statistic.

 Defining the normalized fading gains as in \cite{ZhengTse2003}
\begin{equation}
\label{norm_fad} \alpha_b = -\frac{\log \gamma_b}{\log ({\rm SNR})},
\end{equation}
we have the following result.
\begin{proposition}
\label{prop:asym:alpha} Assume $\gamma_b$ is a random variable with
distribution in (\ref{pdf_gamma}). Assume further that $\alpha_b$ is a random
variable as defined in \eqref{norm_fad}. In this case, the joint distribution
of $\boldalpha = (\alpha_1, \ldots, \alpha_B)$ has the following asymptotic
behavior
\begin{equation}
\label{asymp_alpha} f_{\boldalpha}(\boldalpha) \doteq \left\{
\begin{array}{ll}
\snr^{-m \sum_{b=1}^B \alpha_b}, &\boldalpha \in \field{R}^B_+ \\
0, &\rm{otherwise}.
\end{array}
\right.
\end{equation}
\end{proposition}
\begin{proof}
From (\ref{norm_fad}), $\gamma_b = {\rm SNR}^{-\alpha_b}$. Therefore, the pdf
of $\alpha_b$ is
\begin{eqnarray}
f_{\alpha_b}(\alpha_b) &=& f_{\gamma_b} \left({\rm SNR}^{-\alpha_b}\right) \left|\frac{d \gamma_b}{d \alpha_b}\right| \notag\\
&=& \frac{m^m {\rm SNR}^{-(m-1)\alpha_b} \exp(-m {\rm SNR}^{-\alpha_b})}{\Gamma(m)}{\rm SNR}^{-\alpha_b}\log {\rm SNR} \notag\\
&=& \frac{m^m}{\Gamma(m)} {\rm SNR}^{-m \alpha_b}\exp(-m {\rm SNR}^{-\alpha_b}) \log({\rm SNR}).
\end{eqnarray}
The joint distribution of the vector $\boldalpha$ is then
\begin{equation}
f_{\boldalpha}(\boldalpha) = \left(\frac{m^m \log({\rm
SNR})}{\Gamma(m)}\right)^B {\rm SNR}^{-m\sum_{b=1}^B \alpha_b}
\exp\left(-m\sum_{b=1}^B {\rm SNR}^{-\alpha_b}\right).
\end{equation}
It can easily be seen that
\begin{equation}
\lim_{\snr \rightarrow \infty}
\frac{\log(f_{\boldalpha}(\boldalpha))}{\log(\snr)} = \left\{
\begin{array}{ll}
-m \sum_{b=1}^B \alpha_b, &\boldalpha \in \field{R}^B_+ \\
0, &\rm{otherwise}.
\end{array}
\right.
\end{equation}
Therefore, $f_{\boldalpha}(\boldalpha)$ follows the asymptotic behavior in
(\ref{asymp_alpha}).
\end{proof}

Consider random codes of rate $R$ and block length $L = L({\rm SNR})$ over a
signal set of size $2^M$ such that
\begin{equation}
\lambda = \lim_{{\rm SNR}\rightarrow \infty} \frac{L({\rm SNR})}{\log({\rm
SNR})}.
\end{equation}
Assume the codewords of the code are given by $\mathbf{X}(i), i = 0, \ldots, 2^{BLR}-1$.
Following the analysis in \cite{FabregasCaire2006}, the average pairwise error
probability between $\mathbf{X}(0)$ and $\mathbf{X}(1)$ for a given channel realization
$\mathbf{h}$ is given by
\begin{equation}
\overline{P(\mathbf{X}(0) \to \mathbf{X}(1)|\mathbf{h})} \leq \prod_{b=1}^B
\beta_b^L,
\end{equation}
where $\beta_b$ is the Bhattacharrya coefficient
\begin{equation}
\beta_b = 2^{-2M}\sum_{x\in \mathcal{X}} \sum_{x' \in \mathcal{X}}
\exp\left(-\frac{\snr}{4}\gamma_b|x-x'|^2\right).
\end{equation}
The union bound of the word error probability for a given fading coefficient is
obtained by summing over the pairwise error probability of $2^{BLR}-1$
codewords $\mathbf{X}(i), i = 1, \ldots, 2^{BLR}-1$. Noting that $\gamma_b =
\snr^{1-\alpha_b}$, we obtain
\begin{align}
&\overline{P_e(\snr|\mathbf{h})}  \notag\\
&\leq\exp\left(-BLM \log(2)
\left[1-\frac{R}{M} -\frac{1}{BM}\sum_{b=1}^B
\log_2\left(1+ 2^{-M}\sum_{x \neq x'}  e^{-\frac{1}{4}|x-x'|^2 \snr^{1-\alpha_b}}\right)\right]\right)\\
\label{eq:WER:cond:h}
& = \exp(-BLM \log(2)G(\snr, \boldalpha)).
\end{align}
Using \eqref{eq:WER:cond:h} and the fact that $\overline{P_e(\snr|\mathbf{h})}
\leq 1$, the average error probability is given by
\begin{equation}
\overline{P_e(\snr)} \leq \int_{\boldalpha} \min\{1,\exp(-BLM \log(2)G(\snr,
\boldalpha))\} f_{\boldalpha}(\boldalpha) d\boldalpha.
\end{equation}
Now, from Proposition \ref{prop:asym:alpha}, we have
\begin{equation}
\label{eq:fir:asym:WER}\overline{P_e(\snr)} \dotleq \int_{\boldalpha \in
\field{R}^N_+} \snr^{-m\sum_{b=1}^B \alpha_b}\min\{1,\exp(-BLM \log(2)G(\snr,
\boldalpha))\} d\boldalpha.
\end{equation}
Noting that
\begin{equation}
\lim_{\snr \to \infty} \log_2\left(1+2^{-M}\sum_{x \neq x'}
e^{-\frac{1}{4}|x-x'|^2 \snr^{1-\alpha_b}}\right) = \left\{\begin{array}{ll}
0, &\alpha_b <1 \\
M, &\alpha_b > 1,
\end{array}\right.
\end{equation}
we can replace $G(\snr, \boldalpha)$ in \eqref{eq:fir:asym:WER} by
\begin{equation}
\tilde{G}_{\epsilon} (\boldalpha)= 1 - \frac{R}{M}- \frac{1}{B}\sum_{b=1}^B
\openone\{\alpha_n \geq 1-\epsilon\}
\end{equation}
for any $\epsilon >0$. Therefore, by defining
\begin{equation} \mathcal{B}_{\epsilon} =\left\{\boldalpha:
\tilde{G}_{\epsilon}(\boldalpha) \leq 0\right\}
\end{equation}
and $\mathcal{B}_{\epsilon}^c$ as the complement of $\mathcal{B}_{\epsilon}$,
we can write \eqref{eq:fir:asym:WER} as
\begin{eqnarray}
\label{eq:asym:WER} P_e({\rm SNR}) \dotleq &{}&\int_{\mathcal{B}_{\epsilon} \cap \field{R}_+^B} {\rm
SNR}^{-m\sum_{b=1}^B \alpha_b} d\boldalpha \notag\\
&+&\int_{\mathcal{B}_{\epsilon}^c \cap \field{R}_+^B} \exp\left(-\log({\rm SNR})\left(m \sum_{b=1}^B \alpha_b +
B\lambda M \log(2) \tilde{G}_{\epsilon}(\boldalpha)\right)\right)d\boldalpha.
\end{eqnarray}

By applying the Varadhan's lemma, the SNR-exponents of the first and second
term in \eqref{eq:asym:WER} are given by
\begin{align*}
&\inf_{\boldalpha \in \mathcal{B}_{\epsilon}\cap \field{R}_+^B} \left\{m\sum_{b=1}^B \alpha_b\right\} \;\;\;\; {\rm
and} \;\;\;\inf_{\boldalpha \in \mathcal{B}_{\epsilon}^c \cap \field{R}_+^B}\left\{m \sum_{b=1}^B \alpha_b + B \lambda
M \log(2) \tilde{G}_{\epsilon}(\boldalpha)\right\},
\end{align*}
respectively. Therefore, the SNR-exponent of the word error probability is
given by
\begin{equation}
\label{diversity_achieve} d^{(r)}(R) \geq \sup_{\epsilon >0}
\min\left\{\inf_{\boldalpha \in \mathcal{B}_{\epsilon} \cap \field{R}_+^B}
\left\{m\sum_{b=1}^B \alpha_b\right\}, \inf_{\boldalpha \in
\mathcal{B}_{\epsilon}^c \cap \field{R}_+^B}\left\{m \sum_{b=1}^B \alpha_b + B
\lambda M \log(2) \tilde{G}_{\epsilon}(\boldalpha)\right\}\right\}.
\end{equation}
For the first infimum in (\ref{diversity_achieve}), it can be shown that
\begin{equation}
\inf_{\boldalpha \in \mathcal{B}_{\epsilon} \cap \field{R}_+^B}
\left\{m\sum_{b=1}^B \alpha_b\right\} = m (1- \epsilon)\left\lceil
B\left(1-\frac{R}{M}\right)\right\rceil.
\end{equation}
The infimum is attained when $\left\lceil
B\left(1-\frac{R}{M}\right)\right\rceil$ entries of $\boldalpha$ are
$1-\epsilon$, and the other entries are zero.\\
The second infimum in (\ref{diversity_achieve}) can be rewritten as
\begin{equation}
\label{sec_inf} B \lambda M \log(2)\left(1-\frac{R}{M}\right) +
\inf_{\boldalpha \in \mathcal{B}_{\epsilon}^c \cap \field{R}_+^B}
\left\{\sum_{b=1}^B m \alpha_b - \lambda M \log(2)\openone \{\alpha_b \geq 1-
\epsilon\}\right\}.
\end{equation}
We consider two cases. If $0 \leq \lambda M \log(2) < m$, the infimum in
(\ref{sec_inf}) is zero and achieved when $\boldalpha = \mathbf{0}$. Therefore,
the second infimum in (\ref{diversity_achieve}) is given by
\begin{equation}
B \lambda M \log(2)\left(1- \frac{R}{M}\right).
\end{equation}
If $\lambda M \log(2) \geq m$, the infimum in (\ref{sec_inf}) is given by
\begin{equation}
(m(1- \epsilon) - \lambda M \log(2)) \left\lfloor B\left(1-
\frac{R}{M}\right)\right\rfloor.
\end{equation}
The infimum is attained when $\left\lfloor B\left(1-
\frac{R}{M}\right)\right\rfloor$ entries of $\boldalpha$ are $1-\epsilon$, and
the other entries are zero. Hence, the second infimum in
(\ref{diversity_achieve}) is given by
\begin{equation}
\label{sec_inf_result} B\lambda M \log(2) \left(1-\frac{R}{M}\right) + (m(1-
\epsilon) - \lambda M \log(2)) \left\lfloor B\left(1-
\frac{R}{M}\right)\right\rfloor.
\end{equation}
By collecting the results, and noting that the supremum in
(\ref{diversity_achieve}) is attained when $\epsilon \downarrow 0$, we obtain
the lower bound for the SNR-exponent as in (\ref{eq:diversity_achieve}).
\newpage
\section{Proof of Theorem \ref{diversity_achieve_theorem}}
\label{app:diversity:optimal} Clearly $P_{\rm out}(\snr, R) \geq P^{\ell}_{\rm out}(\snr, R)$, and therefore,
\begin{equation}
\label{proof_theorem} d^{\star}_B(R) \leq m d_B(R)
\end{equation}
follows from Proposition \ref{prop_converse}. In addition, by letting $L(\snr)
\rightarrow \infty$, it follows from Proposition \ref{prop_achieve} that the
SNR-exponent $md_B(R)$ is achievable using random codes for all $R, M$ such
that $B\left(1- \frac{R}{M}\right)$ is not an integer. \hfill\QED

The theorem can also be proved using the SNR-normalized fading coefficients
$\alpha_b \triangleq -\frac{\log(\gamma_b)}{\log(\snr)}$ introduced in
\cite{ZhengTse2003}. The proof given in \cite{FabregasCaire2006} for the
Rayleigh fading case ($m=1$) shows that the asymptotic behavior of the joint
pdf of these coefficients is $f_{\boldalpha}(\boldalpha) \doteq \snr^{-
\sum_{b=1}^B \alpha_b}$ and thus
\begin{equation}
\label{eq:proof7:1} d^\star(R) \leq \inf_{\mathcal{B}_1} \left\{\sum_{b=1}^B
\alpha_b\right\} = d_B(R)
\end{equation}
and
\begin{equation}
\label{eq:proof7:2} d^\star(R) \geq \inf_{\mathcal{B}_2} \left\{\sum_{b=1}^B
\alpha_b\right\} = d_B(R),
\end{equation}
whenever $B\left(1- \frac{R}{M}\right)$ is not an integer, for some suitably
defined sets $\mathcal{B}_1,\mathcal{B}_2$ (see \cite{FabregasCaire2006} for
details). In Proposition \ref{prop:asym:alpha}, it is shown that for
Nakagami-$m$ distributions the asymptotic behavior of the joint pdf of these
coefficients behaves as $f_{\boldalpha}(\boldalpha) \doteq \snr^{-m
\sum_{b=1}^B \alpha_b}$. In this case, the constant $m$ factors out from the
infimums in \eqref{eq:proof7:1} and \eqref{eq:proof7:2} and automatically leads
to the desired result. While this proof is shorter, Proposition
\ref{prop_achieve} provides the extension of the finite block length results of
\cite{FabregasCaire2006}, which illustrates the impact of $m$ in the random
SNR-exponent $d^{(r)}(R)$.

\newpage
\newpage
\newpage
\newpage
\newpage
\nocite{CoverThomas2006}
%

\begin{thebibliography}{10}
\providecommand{\url}[1]{#1}
\csname url@rmstyle\endcsname
\providecommand{\newblock}{\relax}
\providecommand{\bibinfo}[2]{#2}
\providecommand\BIBentrySTDinterwordspacing{\spaceskip=0pt\relax}
\providecommand\BIBentryALTinterwordstretchfactor{4}
\providecommand\BIBentryALTinterwordspacing{\spaceskip=\fontdimen2\font plus
\BIBentryALTinterwordstretchfactor\fontdimen3\font minus
  \fontdimen4\font\relax}
\providecommand\BIBforeignlanguage[2]{{%
\expandafter\ifx\csname l@#1\endcsname\relax
\typeout{** WARNING: IEEEtran.bst: No hyphenation pattern has been}%
\typeout{** loaded for the language `#1'. Using the pattern for}%
\typeout{** the default language instead.}%
\else
\language=\csname l@#1\endcsname
\fi
#2}}

\bibitem{OzarowShamaiWyner1994}
L.~H. Ozarow, S.~Shamai, and A.~D. Wyner, ``Information theoretic
  considerations for cellular mobile radio,'' \emph{IEEE Trans. Veh. Tech.},
  vol.~43, no.~2, pp. 359--378, May 1994.

\bibitem{BiglieriProakisShamai1998}
E.~Biglieri, J.~Proakis, and S.~Shamai, ``Fading channels: Informatic-theoretic
  and communications aspects,'' \emph{IEEE Trans. Inf. Theory}, vol.~44, no.~6,
  pp. 2619--2692, Oct. 1998.

\bibitem{VerduHan1994}
S.~Verd\'{u} and T.~S. Han, ``A general formula for shannon capacity,''
  \emph{IEEE Trans. Inf. Theory}, vol.~40, no.~4, pp. 1147--1157, Jul. 1994.

\bibitem{CaireTariccoBiglieri1999}
G.~Caire, G.~Taricco, and E.~Biglieri, ``Optimal power control over fading
  channels,'' \emph{IEEE Trans. Inf. Theory}, vol.~45, no.~5, pp. 1468--1489,
  Jul. 2001.

\bibitem{Malkamaki1998}
E.~Malkam\"{a}ki, ``Performance of error control over block fading channels
  with {ARQ} applications,'' Ph.D. dissertation, Helsinki Univ. Technology,
  Helsinki, Finland, 1998.

\bibitem{MalkamakiLeib1999}
E.~Malkam\"{a}ki and H.~Leib, ``Coded diversity on block-fading channels,''
  \emph{IEEE Trans. Inf. Theory}, vol.~45, no.~2, pp. 771--781, Mar. 1999.

\bibitem{KnoppHumblet2000}
R.~Knopp and P.~A. Humblet, ``On coding for block fading channels,'' \emph{IEEE
  Trans. Inf. Theory}, vol.~46, no.~1, pp. 189--205, Jan. 2000.

\bibitem{FabregasCaire2006}
A.~{Guill\'{e}n i F\`{a}bregas} and G.~Caire, ``Coded modulation in the
  block-fading channel: Coding theorems and code construction,'' \emph{IEEE
  Trans. Inf. Theory}, vol.~52, no.~1, pp. 91--114, Jan. 2006.

\bibitem{Baccarelli1999}
E.~Baccarelli, ``Asymptotic tight bounds on the capacity and outage probability
  for {QAM} transmission over {R}ayleigh-faded data channels with {CSI},''
  \emph{IEEE Trans. Commun.}, vol.~47, no.~9, pp. 1273--1277, Sep. 1999.

\bibitem{BaccarelliFasano2000}
E.~Baccarelli and A.~Fasano, ``Some simple bounds on the symmetric capacity and
  outage probability for {QAM} wireless channels with {R}ice and {N}akagami
  fadings,'' \emph{IEEE Trans. Veh. Tech.}, vol.~18, no.~3, pp. 361--368, Mar.
  2000.

\bibitem{Nakagami1960}
M.~Nakagami, ``The $m$-distribution - a general formula of intensity
  distribution of rapid fading,'' in \emph{Statistical Methods in Radio Wave
  Propagation}, W.~G. Hoffman, Ed.\hskip 1em plus 0.5em minus 0.4em\relax
  Oxford: Pergamon Press, 1960, pp. 3--36.

\bibitem{SimonAlouini2005}
M.~K. Simon and M.~S. Alouini, \emph{Digital Communications over Fading
  Channels}, 2nd~ed.\hskip 1em plus 0.5em minus 0.4em\relax John Wiley and
  Sons, 2004.

\bibitem{AbramowitzStegun1964}
M.~Abramowitz and I.~A. Stegun, \emph{Handbook of Mathematical Functions with
  Formulas, Graphs, and Mathematical Tables}.\hskip 1em plus 0.5em minus
  0.4em\relax New York: Dover, 1964.

\bibitem{ProakisManolakis1992}
J.~G. Proakis and D.~G. Manolakis, \emph{Digital signal processing :
  principles, algorithms, and applications}, 2nd~ed.\hskip 1em plus 0.5em minus
  0.4em\relax New York : Macmillan, 1992.

\bibitem{ZhengTse2003}
L.~Zheng and D.~N. Tse, ``Diversity and multiplexing: A fundamental tradeoff in
  multiple-antenna channels,'' \emph{IEEE Trans. Inf. Theory}, vol.~49, no.~5,
  pp. 1073--1096, May. 2003.

\bibitem{CoverThomas2006}
T.~M. Cover and J.~A. Thomas, \emph{Elements of Information Theory},
  2nd~ed.\hskip 1em plus 0.5em minus 0.4em\relax John Wiley and Sons, 2006.

\end{thebibliography}

\end{document}